\RequirePackage[mathlines]{lineno}
\documentclass[amsmath,amssymb,twocolumn,prd,floatfix,showpacs, nofootinbib]{revtex4-2}
\usepackage{overpic,graphicx}
\usepackage{dcolumn}
\usepackage{bm}
\usepackage{rotating}
\usepackage{subfigure}
\usepackage{color}
\usepackage[bookmarksnumbered, pdfstartview=FitH,colorlinks,urlcolor=blue, citecolor=blue,linkcolor=blue,] {hyperref}
\usepackage{amsmath}
\usepackage{lineno}
\usepackage{multirow}\usepackage{csquotes}
\usepackage{makecell}%2020.12.10, make the text to be center/right/left
\usepackage{etoolbox}
\makeatletter
\patchcmd{\doauthor}{\unskip}{}{}{}
\makeatother
\newcommand{\BESIIIorcid}[1]{\href{https://orcid.org/#1}{\hspace*{0.1em}\raisebox{-0.45ex}{\includegraphics[width=1em]{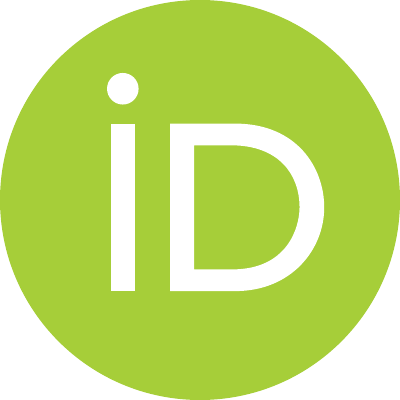}}}}

\newcommand{\dstoksk}{D_{s}^{-} \to K_{S}^{0}K^{-}}

\newcommand{\dstokpipi}{D^{-}_{s} \to K^{-}\pi^{+}\pi^{-}}

\newcommand{\ksk}{K_{S}^{0}K^{-}}
\newcommand{\kkpi}{K^{+}K^{-}\pi^{-}}
\newcommand{\kkpipiz}{K^{+}K^{-}\pi^{-}\pi^{0}}
\newcommand{\kskpipi}{K_{S}^{0}K^{-}\pi^{+}\pi^{-}}
\newcommand{\kskpipim}{K_{S}^{0}K^{+}\pi^{-}\pi^{-}}
\newcommand{\pipipi}{\pi^{+}\pi^{-}\pi^{-}}
\newcommand{\pieta}{\pi^{-}\eta}
\newcommand{\pipizeta}{\pi^{-}\pi^{0}\eta}
\newcommand{\pietapgam}{\pi^{-}\eta'_{\pi^{+}\pi^{-}\eta}}
\newcommand{\pietaprho}{\pi^{-}\eta'_{\gamma \rho^{0}}}
\newcommand{\kpipi}{K^{-}\pi^{+}\pi^{-}}

\newcommand{\dstotaunuz}{D_{s}^{+} \to \tau^{+} \nu_{\tau}}

\let\oldequation\equation
\let\oldendequation\endequation
\renewenvironment{equation}
{\linenomathNonumbers\oldequation}
{\oldendequation\endlinenomath}

\begin{document}
%	\linenumbers
	
	\title{\bf \boldmath
		Updated measurement of the branching fraction of $D_s^+ \to \tau^+ \nu_{\tau}$ 
	}

	\author{%% Saved at => 2025-09-10
M.~Ablikim$^{1}$\BESIIIorcid{0000-0002-3935-619X},
M.~N.~Achasov$^{4,b}$\BESIIIorcid{0000-0002-9400-8622},
P.~Adlarson$^{81}$\BESIIIorcid{0000-0001-6280-3851},
X.~C.~Ai$^{86}$\BESIIIorcid{0000-0003-3856-2415},
C.~S.~Akondi$^{31A,31B}$\BESIIIorcid{0000-0001-6303-5217},
R.~Aliberti$^{39}$\BESIIIorcid{0000-0003-3500-4012},
A.~Amoroso$^{80A,80C}$\BESIIIorcid{0000-0002-3095-8610},
Q.~An$^{77,64,\dagger}$,
Y.~H.~An$^{86}$\BESIIIorcid{0009-0008-3419-0849},
Y.~Bai$^{62}$\BESIIIorcid{0000-0001-6593-5665}
% ...
\\
\vspace{0.2cm}
(BESIII Collaboration)\\
\vspace{0.2cm} {\it
$^{1}$ Institute of High Energy Physics, Beijing 100049, People's Republic of China\\
% ...
$^{4}$ Budker Institute of Nuclear Physics SB RAS (BINP), Novosibirsk 630090, Russia\\
% ...
$^{31}$ INFN Sezione di Ferrara, (A)INFN Sezione di Ferrara, I-44122, Ferrara, Italy; (B)University of Ferrara, I-44122, Ferrara, Italy\\
% ...
$^{39}$ Johannes Gutenberg University of Mainz, Johann-Joachim-Becher-Weg 45, D-55099 Mainz, Germany\\
% ...
$^{62}$ Southeast University, Nanjing 211100, People's Republic of China\\
% ...
$^{64}$ State Key Laboratory of Particle Detection and Electronics, Beijing 100049, Hefei 230026, People's Republic of China\\
% ...
$^{77}$ University of Science and Technology of China, Hefei 230026, People's Republic of China\\
% ...
$^{80}$ University of Turin and INFN, (A)University of Turin, I-10125, Turin, Italy; (B)University of Eastern Piedmont, I-15121, Alessandria, Italy; (C)INFN, I-10125, Turin, Italy\\
$^{81}$ Uppsala University, Box 516, SE-75120 Uppsala, Sweden\\
% ...
$^{86}$ Zhengzhou University, Zhengzhou 450001, People's Republic of China\\

\vspace{0.2cm}
$^{\dagger}$ Deceased\\
% ...
$^{b}$ Also at the Novosibirsk State University, Novosibirsk, 630090, Russia\\
% ...

}
%% ends here %%
}

\begin{abstract}
	Using a dataset of 7.33~fb$^{-1}$ of $e^+ e^-$ annihilation data collected with the BESIII detector at center-of-mass energies from 4.128 to 4.226~GeV, we report an updated measurement of the branching fraction of $D_s^+ \to \tau^+ \nu_{\tau}$ via four $\tau^+$ decay modes: $\tau^+ \to e^+ \nu_e \bar{\nu}_{\tau}$, $\mu^+ \nu_{\mu} \bar{\nu}_{\tau}$, $\pi^+\bar{\nu}_{\tau}$, and $\pi^+ \pi^{0} \bar{\nu}_{\tau}$.
	The branching fraction is determined to be $\mathcal{B}({D_s^+ \to \tau^+ \nu_{\tau}}) = (5.37 \pm 0.08_{\rm stat} \pm  0.06_{\rm syst})$\%. 
	The product of the modulus of the Cabibbo-Kobayashi-Maskawa matrix element $|V_{cs}|$ and the $D_s^+$ decay constant $f_{D_s^+}$ 
	is measured to be $f_{D_s^+} |V_{cs}| = (248.2 \pm 1.9_{\rm stat} \pm 1.4_{\rm syst} \pm 0.6_{\rm input} \pm 0.8_{\rm EM})$~MeV. Both the branching fraction and the product $f_{D_s^+} |V_{cs}|$ are the most precise results yet obtained.
	Then, taking $f_{D_s^+}$ from lattice quantum chromodynamics calculations results in $|V_{cs}| = 0.993 \pm 0.008_{\rm stat} \pm 0.006_{\rm syst} \pm 0.003_{\rm input} \pm 0.003_{\rm EM}$. 
	Conversely, one finds $f_{D_s^+} = (255.0 \pm 1.9_{\rm stat} \pm 1.4_{\rm syst} \pm 0.6_{\rm input} \pm 0.8_{\rm EM})$~MeV when taking $|V_{cs}|$ from the CKMfitter group as an input.
	Combining with the world average value of $D_s^+ \to \mu^+ \nu_{\mu}$, the ratio of the branching fractions between $D_s^+ \to \tau^+ \nu_{\tau}$ and $D_s^+ \to \mu^+ \nu_{\mu}$ is estimated to be $10.04 \pm 0.29$, 
	which is consistent with the Standard Model prediction of lepton flavor universality.

\end{abstract}		

\maketitle

\oddsidemargin  -0.2cm
\evensidemargin -0.2cm
\section{INTRODUCTION}

\begin{figure}[hbpt]
	\centering
	\includegraphics[width=0.3\textwidth]{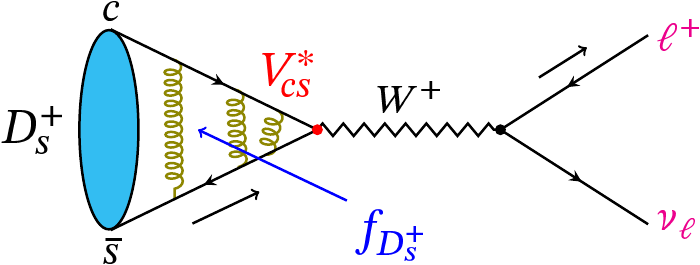}
	\caption{The Feynman diagram of $D_s^+ \to \ell^+ \nu_{\ell}$ ($\ell=e, \mu, \tau$) decays.}
	\label{feynman_leptonic_dsp}
\end{figure}

In the Standard Model (SM), the leptonic $D_s^+$ decays proceed via the annihilation of the initial $c$ and $\bar{s}$ quarks into a virtual $W^+$ boson, 
which then materializes as an $\ell^+ \nu_{\ell}$ pair ($\ell=e, \mu, \tau$). Figure~\ref{feynman_leptonic_dsp} shows the Feynman diagram of a $D_s^+ \to \ell^+ \nu_{\ell}$ decay. 
Throughout this paper, charge-conjugate channels are implied. 
Due to the zero spin of the $D_s^+$ meson and the left-handed neutrino, chirality considerations require the charged lepton to be right-handed. 
The $D_s^+ \to e^+ \nu_e$ decay, which has a branching fraction (BF) predicted by the SM to be approximately $10^{-8}$, is highly helicity suppressed and has not yet been observed. 
However, the helicity suppression is largely avoided for the $\tau^+$ case, since the mass of $\tau^+$ lepton is close to that of $D_s^+$, albeit with significant phase-space suppression.

According to the SM, the partial decay width for $D_s^+ \to \ell^+ \nu_{\ell}$ can be written as~\cite{decayratedstotaunu}   

\begin{equation}
	\label{decayratedstolnu_draft}
	\begin{split}
		\Gamma_{D_s^+ \to \ell^+ \nu_\ell} &= \Gamma_{D_s^+ \to \ell^+ \nu_\ell}^{(0)} \left[ 1 + \frac{\alpha}{\pi} C_p \right] \\
		&= \frac{G_F^2 f_{D_s^+}^2 m_{D_s^+}^3}{8\pi} \vert V_{cs} \vert^2 \mu_\ell^2 (1 - \mu_\ell^2)^2 \left[ 1 + \frac{\alpha}{\pi} C_p \right],
	\end{split}
\end{equation}
where $G_{F}$ is the Fermi coupling constant, 
$f_{D_s^+}$ is the $D_s^+$ decay constant encapsulating the non-perturbative strong interactions that bind the initial $c$ and $\bar{s}$ quarks within the $D_s^+$ meson, $|V_{cs}|$ is the modulus of the $c \to s$ Cabibbo-Kobayashi-Maskawa (CKM) matrix element, 
$m_{D_s^+}$ is the mass of $D_s^+$ meson, $\mu_{\ell}$ is the ratio of the $\ell^+$ lepton mass to the $D_s^+$ meson mass, and $[1+(\alpha/\pi)C_{p}]$ represents the radiative correction term~\cite{pdg}.  
Therefore, an accurate measurement of the BF for $D_{s}^{+} \to \ell^+ \nu_\ell$ enables a clean determination of the product $|V_{cs}| f_{D_s^+}$. By substituting the value of $f_{D_s^+}$ from lattice quantum chromodynamics (LQCD) calculations~\cite{fds_flag}, a precise value of $|V_{cs}|$ can be obtained, and by taking the $|V_{cs}|$ from the SM global fit~\cite{pdg} as input, the decay constant $f_{D_s^+}$ can be extracted with high precision.  
Thus, precision studies of leptonic $D_s^+$ decays play a key role in calibrating LQCD calculations and testing the unitarity of the CKM matrix.

The couplings between the three families of leptons and the gauge bosons are expected to be the same in the SM; this is referred to as lepton flavor universality (LFU). 
It demands that the ratios of BFs for leptonic $D_s^+$ decays are only from phase space and helicity suppression. From Eq.~(\ref{decayratedstolnu_draft}), these ratios are $2.35\times 10^{-5}$:~1:~9.75 for $e^+$:~$\mu^+$:~$\tau^+$, with negligible uncertainties of the input masses.
The experimental measurements for the latter two cases align with the SM predictions. 
However, hints of LFU violation in $b\to c\, \ell^{+} \bar{\nu}_{\ell}$ decays have been reported by several experiments~\cite{flv1,flv2,flv3,flv4,flv5}.
Violations might also appear in $c\to s$ transitions, via the interference among different amplitudes~\cite{lfu_dstotaunu} or interactions involving scalar operators~\cite{liying}. 
Consequently, high-precision BF measurements for leptonic $D_s^+$ decays are interesting probes of LFU violation, which can then be used to limit new physics.

Previous studies of $D_s^+ \to \tau^+ \nu_{\tau}$ via different $\tau^+$ decays have been reported by the CLEO~\cite{cleo_dstotaunu_enunu1,cleo_dstotaunu_enunu2,cleodstotaunu2,cleodstotaunu3},
$BABAR$~\cite{babardstotaunu}, Belle~\cite{belldstotaunu}, and BESIII~\cite{bes4009,tauvhajime,tauvyue,tauvhuijing,tauvxiechen,tauvmununu} experiments.
%via different $\tau^+$ decays individually, 
Each collaboration has provided averages where the correlated uncertainties of different results have been considered.  
Recently, the BESIII experiment~\cite{tauvwuxiao} adopted a simultaneous fit method on $D_s^+ \to \tau^+ \nu_{\tau}$ for the different $\tau^+$ decays, based on the data samples taken at center-of-mass energies ($E_{\rm CM}$) from 4.237 to 4.699~GeV with the BESIII detector~\cite{Ablikim:2009aa,luopw_ds}, where the $D_s^\pm$ mesons are produced via $D_s^{*\pm}$ decays.
But the results were limited by the statistics due to the low cross sections of $e^+ e^- \to D_s^{*+} D_s^{*-}$.  
In this paper, we report an improved measurement of $\mathcal{B}({D_s^+ \to \tau^+ \nu_{\tau}})$, by combining four $\tau^+$ decay modes of $\tau^+ \to e^+ \nu_e \bar{\nu}_{\tau}$, $\mu^+ \nu_{\mu} \bar{\nu}_{\tau}$, $\pi^+ \bar{\nu}_{\tau}$, and
$\pi^+ \pi^{0} \bar{\nu}_{\tau}$. 
In contrast, the $e^+e^-$ collision data used in this work, corresponding to a total integrated luminosity of 7.33~fb$^{-1}$, were taken at $E_{\rm CM}=$ 4.128, 4.157, 4.178, 4.189, 4.199, 4.209, 4.219, and around 4.226~GeV~\cite{ref_emc_4230_hajime, ref_lumi_4230_hajime} by the BESIII detector.
At these energies, one makes use of the $e^+ e^- \to D_s^{*+} D_s^{-}$ process, with a higher cross-section than $e^+ e^- \to D_s^{*+} D_s^{*-}$ which was used in the previous BESIII analysis. 
%To be clear, we simply use $\tau_{e}^+$, $\tau_{\mu}^+$, $\tau_{\pi}^+$, and $\tau_{\rho}^+$, to represent the four $\tau^+$ modes in the following text. 

\section{BESIII DETECTOR AND MONTE CARLO SIMULATION}

The BESIII detector~\cite{Ablikim:2009aa} operates at the BEPCII $e^+e^-$ collider~\cite{Yu:IPAC2016-TUYA01}, which delivers symmetric collisions within a center-of-mass energy range of 1.84-4.95~GeV. The machine achieves its maximum luminosity of $1.1\times10^{33}\,\text{cm}^{-2}\text{s}^{-1}$ at $E_{\rm CM}=3.773$~GeV.
BESIII has accumulated extensive datasets at a variety of energies~\cite{Ablikim:2019hff}. The detector is designed with a cylindrical geometry that covers approximately 93\% of the full solid angle.
From the interaction point outward, it consists of a helium-based multilayer drift chamber (MDC), a time-of-flight (TOF) system, and a CsI(Tl) electromagnetic calorimeter (EMC), all surrounded by a superconducting solenoid generating a 1.0~T magnetic field.
The solenoid is enclosed by an octagonal flux-return yoke, which is instrumented with resistive plate chambers serving as muon counters (MUC). For charged particles with a momentum of $1~{\rm GeV}/c$, the tracking resolution is about 0.5\%, and the ${\rm d}E/{\rm d}x$ resolution for electrons from Bhabha scattering events reaches 6\%.
Photon energy is measured by the EMC with a resolution of 2.5\% in the barrel region and 5\% in the end caps at 1~GeV.
The TOF system provides a timing resolution of 68~ps in the barrel and roughly 110~ps in the end-cap region.
Since 2015, the end-cap TOF has been upgraded to a multigap resistive plate chamber (MRPC) system, improving the timing performance to about 60~ps, which applies to approximately 83\% of the dataset used in this analysis~\cite{etof}.

Monte Carlo (MC) simulated samples are generated using a framework based on {\sc geant4}~\cite{geant4}, which provides a detailed modeling of the BESIII detector geometry and its response. These samples are employed to evaluate detection efficiencies and to estimate background contributions. The simulation incorporates effects such as the beam energy spread and initial-state radiation (ISR) in $e^+e^-$ annihilations through the {\sc kkmc} generator~\cite{ref:kkmc}. An inclusive MC dataset, corresponding to approximately 40 times the integrated luminosity of the real data and covering center-of-mass energies from 4.128 to 4.226~GeV, is produced. It includes open-charm processes, ISR production of $\psi(3770)$, $\psi(3686)$, and $J/\psi$, continuum $q\bar q$ $(q=u, d, s)$ interactions, as well as Bhabha, $\mu^+\mu^-$, $\tau^+\tau^-$, and $\gamma\gamma$ events. The generation of open-charm processes directly produced in $e^+e^-$ collisions is performed using the {\sc conexc} package~\cite{ref:conexc}. Known decay modes are simulated with {\sc evtgen}~\cite{ref:evtgen}, where BFs are taken from the Particle Data Group (PDG)~\cite{pdg} whenever available, and otherwise modeled with {\sc lundcharm}~\cite{ref:lundcharm}. Final-state radiation emitted by charged particles is handled by the {\sc photos} package~\cite{photos2}.

\section{Analysis method}
A ``double-tag" (DT) method~\cite{dtmethod1, dtmethod2} is employed to investigate the leptonic $D_s^+$ decay from the events of $e^+ e^- \to D_s^{*+} D_s^- \to \gamma (\pi^0) D_s^+ D_s^-$, in the presence of $D_s^-$ fully reconstructed via several hadronic decay modes as ``single-tag" (ST) side.
Benefiting from this property, the BF of $D_s^+ \to \tau^+ \nu_{\tau}$ can be determined by 
\begin{equation}
	%\mathcal B_{\rm sig}=\frac{N_{\rm DT}}{N_{\rm ST} \, \bar\epsilon_{{\rm sig}} \cdot \mathcal{B}_{\rm sub}}.
	\mathcal \mathcal{B}({D_s^+ \to \tau^+ \nu_{\tau}})=\frac{N_{\rm DT}}{\sum_{i,\alpha} (N_{\rm ST}^{i,\alpha} \cdot \epsilon_{\rm DT}^{i,\alpha}/\epsilon_{\rm ST}^{i,\alpha})}.
	\label{eq1_br_cal}
\end{equation}
Here $N_{\rm ST}^{i,\alpha}$, $\epsilon_{\rm ST}^{i,\alpha}$, and $\epsilon_{\rm DT}^{i,\alpha}$ are the ST yield, ST efficiency, and DT efficiency for the $i^{\rm th}$ tag mode at the $\alpha^{\rm th}$ energy point, respectively. 
The $N_{\rm DT}$ is the total DT yield. The BFs of the intermediate-state decays of $\tau^+$, $K_S^0$, $\pi^0$, $\rho^0$, and $\eta^{(\prime)}$ have been included in the relevant efficiency.

\section{Single Tag $D_s^-$ candidate}
\label{sec_st}

The ST $D^-_s$ candidates are reconstructed from eleven hadronic tag modes: $D_s^-\to\ksk$, $\kkpi$, $\kkpipiz$, $\kskpipi$, $\kskpipim$, $\pipipi$, $\pieta$, $\pipizeta$, $\pietapgam$, $\pietaprho$, and $\kpipi$, followed by the intermediate-state decays $K_S^0 \to \pi^+ \pi^-$, $\pi^0 (\eta)\to \gamma \gamma$, $\eta^{\prime}_{\pi^+ \pi^- \eta} \to \pi^+ \pi^- \eta$, $\eta^{\prime}_{\gamma \rho^0} \to \gamma \rho^0$, and $\rho^0\to\pi^+ \pi^-$.

%good charged tracks selection criteria
Charged tracks detected in the MDC are required to be within a polar angle ($\theta$) range of $|\rm{cos\theta}|<0.93$, where $\theta$ is defined with respect to the $z$-axis, which is the symmetry axis of the MDC.
For charged tracks not originating from $K_S^0$ decays, the distance of closest approach to the interaction point (IP) must be less than 10\,cm along the $z$-axis, $|V_{z}|$, and less than 1\,cm in the transverse plane, $|V_{xy}|$.
Particle identification~(PID) for charged tracks combines measurements of the energy deposited in the MDC and the time of flight in the TOF to form likelihoods $\mathcal{L}(h)~(h=K,\pi)$ for each hadron $h$ hypothesis. 
Charged kaons and pions are identified by requiring the likelihoods for the kaon and pion hypotheses, $\mathcal{L}(K)>\mathcal{L}(\pi)$ and $\mathcal{L}(\pi)>\mathcal{L}(K)$, respectively.

%ks candidate
Each $K_{S}^0$ candidate is reconstructed from two oppositely charged tracks satisfying $|V_{z}|<$ 20~cm. 
The two charged tracks are assigned as $\pi^+\pi^-$ without imposing further PID criteria. They are constrained to originate from a common vertex with a loose fit-quality requirement of $\chi^2 < 100$, and are required to have an invariant mass $M_{\pi^{+}\pi^{-}}$ within $[0.487, 0.511]$~GeV$/c^{2}$.
The decay length of the $K^0_S$ candidate is required to be greater than twice the vertex resolution away from the IP.

%photons
Photon candidates are identified using isolated showers in the EMC. The deposited energy of each shower must be more than 25~MeV in the barrel region, $|\cos \theta|< 0.80$, and more than 50~MeV in the endcap regions, $0.86 <|\cos \theta|< 0.92$.
To exclude showers that originate from charged tracks, the angle subtended by the EMC shower and the position of the closest charged track at the EMC must be greater than 10 degrees as measured from the IP.
To suppress electronic noise and showers unrelated to the event, the difference between the EMC time and the event start time is required to be within [0, 700]\,ns.

%pi0/eta/eta'/rho
Candidates of $\pi^0$ or $\eta$ are formed with pairs of photons with invariant mass $M_{\gamma \gamma}$ in the interval [0.115, 0.150] or [0.50, 0.57]~GeV/$c^2$.
To improve the momentum resolution, the photon pair is kinematically constrained to the $\pi^0$ or $\eta$ nominal mass. For the $\rho^0$ and $\eta^{\prime}$ candidates, the invariant masses of the $\pi^+ \pi^-$, $\pi^+ \pi^- \eta$, and $\gamma \rho^0$ combinations are required to be within the mass intervals of [0.500, 0.970], [0.946, 0.970], and [0.940, 0.976]~GeV/$c^2$, respectively.
In order to suppress the transition pion and soft $\gamma$ from $D^{*}$ decays, a momentum greater than 0.1~GeV/$c$ is required for each pion and $\gamma$ from the $\eta^{\prime}\to \gamma \rho^0$ decay. To avoid the overlap between $\dstoksk$ and $\dstokpipi$, events within $M_{\pi^+ \pi^-}\in [0.480, 0.515]$~GeV/$c^2$ are rejected for the latter mode.

\begin{figure}[htbp] 
	\centering 
%	\begin{minipage}{0.22\textwidth} 
%		\centering 
		\includegraphics[width=0.48\linewidth]{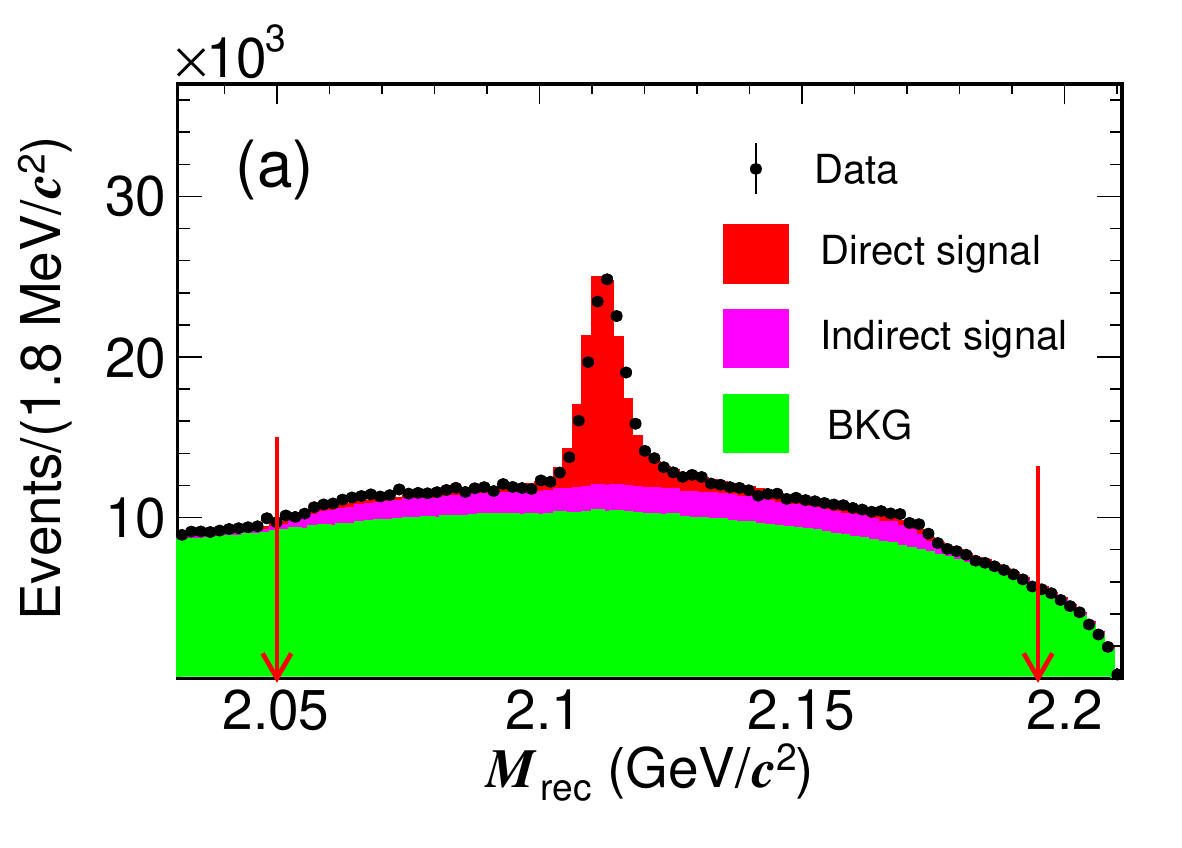} 
%		\vspace{0.2cm} {\small (a)} 
%	\end{minipage} 
%   \hfill 
%   \begin{minipage}{0.22\textwidth} 
%   	\centering 
   	\includegraphics[width=0.48\linewidth]{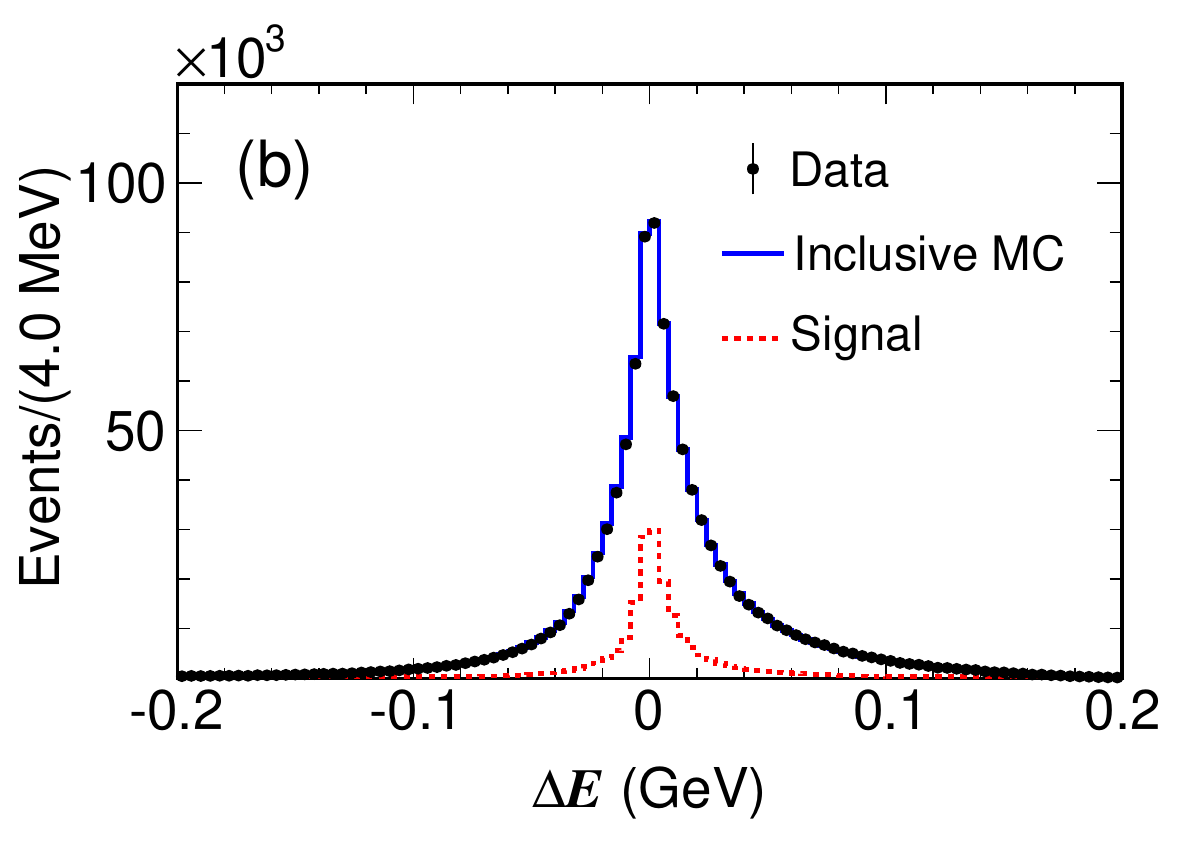} 
%   	\vspace{0.2cm} {\small (b)} 
%   \end{minipage}
 \caption{The distributions of $M_{\rm rec}$ (a) and
 	$\Delta E$ (b) for the
 	$D_s^- \rightarrow K^+K^-\pi^-$ tag mode at
 	$\sqrt{s}=4.178$ GeV in data and MC-simulated samples. 
% 	The dots with error bars are data, 
% 	the solid blue line is the inclusive MC samples normalized to data, 
% 	and the dashed red line is the signal shape scaled to data. 
 	The area between two red arrows indicates the signal region.} 
 \label{fig:STselection}
 \end{figure}

The recoil mass against the ST $D_s^-$,  
\begin{equation}
	M_{\rm rec}^2c^4 = \Big (E_{\rm CM} - \sqrt{|\vec{p}_{\rm ST}|^2c^2+m^2_{D_s^-}c^4}\Big )^2 - |\vec{p}_{\rm ST}|^2c^2,
	\label{eq2_mrec_st}
\end{equation}
is utilized to suppress the non-$D_s^{*+} D_s^-$ background events, where $\vec{p}_{\rm ST}$ is the three-momentum of the reconstructed ST $D_s^-$, 
and $m_{D_s^-}$ is the nominal $D_s^-$ mass.
Since the ST $D_s^-$ candidates either arise from the $e^+ e^-$ collision directly, or indirectly from $D_s^{*-}$ decays~\cite{tauvxiechen}, 
events within a broad range of $M_{\rm rec}$, 
namely with $M_{\rm rec} \in[2.050, 2.195]$~GeV/$c^2$, are retained for further analyses.
If multiple ST $D_s^-$ candidates are present per event, the one with the minimum value of $|M_{\rm rec}-m_{D_s^{*+}}|$ is kept for each tag mode, where $m_{D_s^{*+}}$ is the $D_s^{*+}$ nominal mass~\cite{pdg}. 
Figure~\ref{fig:STselection}(a) shows the $M_{\rm rec}$ distribution for the
  $D_s^- \rightarrow K^+K^-\pi^-$ tag mode at
  $\sqrt{s}=4.178$ GeV as an example.

\begin{figure}[hbpt]
	\centering
	\includegraphics[width=0.50\textwidth]{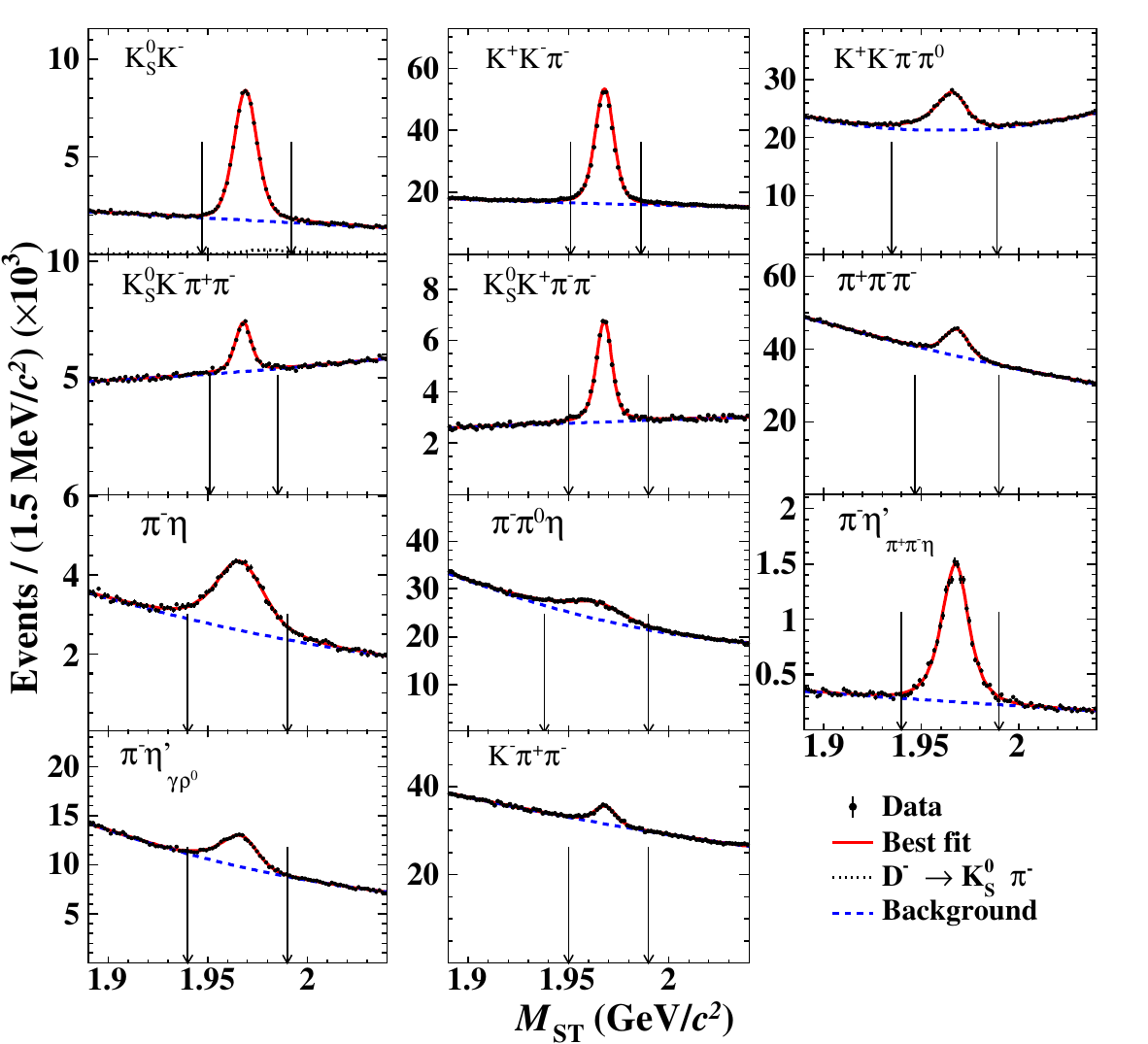}
	\caption{Fits to the invariant mass distributions of the ST $D_s^-$ candidates, $M_{\rm ST}$, at $E_{\rm CM}=$ 4.128-4.226~GeV. The dots with error bars are data, the solid red lines are the fit results, the dashed blue lines are the combinatorial backgrounds and the dotted black line is the $D^-\to K_S^0 \pi^-$ background for the $D_s^- \to K_S^0 K^-$ tag mode. The area between two black arrows denotes the signal region.}
	\label{ST_yield_data_4178}
\end{figure}

The ST yield is derived from a fit to the invariant mass of the reconstructed ST $D_s^-$ candidate ($M_{\rm ST}$). 
In the fit, the signal shape is described by the MC-simulated shape convolved with a Gaussian function, which is employed to compensate for the resolution difference between data and MC simulation. The combinatorial backgrounds are described by a first to third order polynomial.
For the $\dstoksk$ mode, the peaking background $D^- \to K_S^0 \pi^-$ is modeled by the MC-simulated shape with a floating yield. 
Figure~\ref{ST_yield_data_4178} shows the fit results for the data samples at $E_{\rm CM}=4.128-4.226$~GeV.
In each plot, the area between two black arrows is specified as the $M_{\rm ST}$ signal region, corresponding to approximately $\pm 3\sigma$ around the $D_s^-$ mass.
Here, $\sigma$ is the fitted resolution of $M_{\rm ST}$. 
The $\chi^2/$ndf value is also displayed on each plot to demonstrate the goodness of fit, 
where the ndf means the number of degree of freedom. 
The associated ST efficiency is estimated with the inclusive MC sample following the same procedure as for data. 
The $M_{\rm ST}$ mass window, ST yield, and ST efficiency for each mode are listed in Table~\ref{tab_st_dt}. The total ST yield is $(801.7\pm 3.8) \times 10^{3}$ summing over all of the tag modes at all energy points.

\begin{table*}[htbp]
	\caption{The $M_{\rm ST}$ requirements, ST yields ($N_{\rm ST}$), ST efficiencies ($\epsilon_{\rm ST}$) for each tag mode after combining eight energy points, as well as the corresponding DT efficiencies for the four cases:
		$\tau^+_{e}$ ($\epsilon^{e}_{\rm DT}$), 
		$\tau^+_{\mu}$ ($\epsilon^{\mu}_{\rm DT}$), 
		$\tau^+_{\pi}$ ($\epsilon^{\pi}_{\rm DT}$), 
		and $\tau^+_{\rho}$ ($\epsilon^{\rho}_{\rm DT}$). 
		The efficiencies include the BFs of the intermediate-state decays ($\tau^+$, $K_S^{0}$, $\pi^{0}$, $\rho^0$, and $\eta^{(\prime)}$). The uncertainties are statistical only.}
	\centering
	\label{tab_st_dt}
	\setlength{\extrarowheight}{0.2ex}%the space left in the top
	\setlength{\tabcolsep}{3.5pt}% the space between two columns
	\renewcommand{\arraystretch}{1.5}% the space between two rows
	%\begin{tabular}{ l l   r@{} lr@{}  lr@{} lr@{}  lr@{} lr@{} lr@{}l }
	\begin{tabular}{ l l   r@{}l  r@{}l  r@{}l r@{}l  r@{}l  r@{}l }
		\hline\hline
		ST mode 
		&\makecell{$M_{\rm ST}$ (GeV/$c^2$)\\}
		&\multicolumn{2}{c}{$N_{\rm ST} (\times 10^{3})$}
		&\multicolumn{2}{c}{\makecell{$\epsilon_{\rm ST}$(\%)\\}}
		&\multicolumn{2}{c}{\makecell{$\epsilon^{e}_{\rm DT}$(\%)\\}}
		&\multicolumn{2}{c}{\makecell{$\epsilon^{\mu}_{\rm DT}$(\%)\\}}
		&\multicolumn{2}{c}{\makecell{$\epsilon^{\pi}_{\rm DT}$(\%)\\}}
		&\multicolumn{2}{c}{\makecell{$\epsilon^{\rho}_{\rm DT}$(\%)\\}}\\
		\hline
		$D_s^- \to K_{S}^{0}K^{-}$     &$[1.947,1.992]$&$65.7$&$\pm0.4$&$34.43   $&$\pm0.04$&$18.21   $&$\pm0.19$&$8.53   $&$\pm0.36$&$16.12    $&$\pm0.13$&$6.38    $&$\pm0.13$\\
		$D_s^- \to K^{+}K^{-}\pi^{-}$   &$[1.951,1.986]$&$284.3$&$\pm0.9$&$41.00    $&$\pm0.02$&$20.03    $&$\pm0.11$&$9.56    $&$\pm0.23$&$17.35    $&$\pm0.08$&$6.00    $&$\pm0.08$\\
		$D_s^- \to K^{+}K^{-}\pi^{-}\pi^{0}$  &$[1.935,1.989]$&$100.8$&$\pm1.5$&$13.77   $&$\pm0.03$&$7.24   $&$\pm0.07$&$3.58   $&$\pm0.17$&$5.93    $&$\pm0.05$&$1.71    $&$\pm0.05$\\
		$D_s^- \to K_{S}^{0}K^{-}\pi^{+}\pi^{-}$ &$[1.951,1.985]$&$15.7$&$\pm0.4$&$13.28   $&$\pm0.06$ &$6.27   $&$\pm0.17$&$3.12   $&$\pm0.31$&$5.27    $&$\pm0.12$&$2.03    $&$\pm0.12$\\
		$D_s^- \to K_{S}^{0}K^{+}\pi^{-}\pi^{-}$   &$[1.950,1.990]$&$29.1$&$\pm0.3$&$14.39    $&$\pm0.03$&$7.09    $&$\pm0.13$&$3.29    $&$\pm0.25$&$6.18    $&$\pm0.09$&$2.03    $&$\pm0.07$\\
		$D_s^- \to \pi^{+}\pi^{-}\pi^{-}$   &$[1.947,1.990]$&$81.6$&$\pm1.4$&$56.36   $&$\pm0.13$&$28.55   $&$\pm0.28$&$13.89   $&$\pm0.77$&$27.61    $&$\pm0.21$&$10.91    $&$\pm0.30$\\
		$D_s^- \to \pi^{-}\eta$ &$[1.940,1.990]$&$34.6$&$\pm0.7$&$16.49   $&$\pm0.04$&$9.19    $&$\pm0.12$&$4.21   $&$\pm0.28$&$8.59    $&$\pm0.08$&$3.69    $&$\pm0.09$\\
		$D_s^- \to \pi^{-}\pi^{0}\eta$  &$[1.938,1.990]$&$84.0$&$\pm2.4$&$7.06   $&$\pm0.03$&$4.23    $&$\pm0.03$&$1.99    $&$\pm0.10$&$4.29    $&$\pm0.03$&$1.64    $&$\pm0.03$\\
		$D_s^- \to \pi^{-}\eta'_{\pi^{+}\pi^{-}\eta}$  &$[1.940,1.990]$&$15.9$&$\pm0.2$&$3.33    $&$\pm0.01$&$1.81    $&$\pm0.03$&$0.81    $&$\pm0.07$&$1.66    $&$\pm0.03$&$0.66    $&$\pm0.02$\\
		$D_s^- \to \pi^{-}\eta'_{\gamma \rho^{0}}$   &$[1.940,1.990]$&$51.3$&$\pm1.0$&$9.72   $&$\pm0.03$&$5.24   $&$\pm0.06$&$2.52   $&$\pm0.15$&$5.23    $&$\pm0.05$&$1.94    $&$\pm0.05$\\
		$D_s^- \to K^{-}\pi^{+}\pi^{-}$  &$[1.950,1.990]$&$38.7$&$\pm1.2$&$49.08   $&$\pm0.21$&$23.92    $&$\pm0.39$&$11.42   $&$\pm0.90$&$22.60    $&$\pm0.30$&$9.54    $&$\pm0.50$\\
		\hline\hline
	\end{tabular}
	
\end{table*}

\section{Transition $\gamma(\pi^0)$ candidate}
In the presence of the best ST $D_s^-$ candidate per event, the transition $\gamma(\pi^0)$ candidate from the $D_s^{*+}$ decay is reconstructed with the same criteria from Sec.~\ref{sec_st}.
Then the kinematic variable $\Delta E$ is employed, defined by
\begin{equation}
	\label{eq:dele}
	\Delta E = E_{\rm CM}-E_{\rm ST}-E_{\rm miss}^{\prime}-E_{\gamma(\pi^0)},
\end{equation}
\noindent where $E_{\rm miss}^{\prime}$ represents the corresponding missing energy. 
In order to improve the resolution, the energy of ST $D_s^-$ is calculated as $E_{\rm ST} = \sqrt{|\vec{p}_{\rm ST}|^2c^2+m_{D_s^-}^2c^4}$. 
The ST $D_s^-$ and transition $\gamma(\pi^0)$ recoil against the $D_s^+$ candidate. 
Thus, the missing energy is written as $E^{\prime}_{\rm miss} = \sqrt{|\vec{p}^{\;\prime}_{\rm miss}|^2 c^2+m_{D_s^+}^2 c^4}$, along with missing three-momentum $\vec{p}^{\;\prime}_{\rm miss} = -\vec{p}_{\rm ST}-\vec{p}_{\gamma(\pi^0)}$. A very loose requirement of $\Delta E\in[-0.2, 0.2]$~GeV is applied with negligible signal efficiency loss.
In case of multiple $\gamma$ and $\pi^0$ candidates, the one with the minimum $|\Delta E|$ is kept for the further analysis. 
Figure~\ref{fig:STselection}(b) shows the $\Delta E$ distribution for the
  $D_s^- \rightarrow K^+K^-\pi^-$ tag mode at
  $\sqrt{s}=4.178$ GeV as an example.
%The distributions of the $\Delta E$ is shown in Fig.~\ref{fig:STselection}.

\section{Signal $D^+_s \to \tau^+\nu_\tau$ decay}
%%%%%%%%%%%%%%%%%%%%%%%%%%%%%%%%%%%%%%%%%%%%%%%%%%%%%%%%%%%%%%%%%%%%%%%%%%%%%%%%%%%%%%%%%%%%%%%%%%%%%%%%%%%%%
In the recoil side against the ST $D_s^-$ and the transition $\gamma(\pi^0)$ from $D_s^{*+}$ decay, the signal $D_s^+ \to \tau^+ \nu_\tau$ is investigated for the four cases:
$\tau^+ \to e^+ \nu_e \bar{\nu}_{\tau}$, 
$\mu^+ \nu_{\mu} \bar{\nu}_{\tau}$, 
$\pi^+ \bar{\nu}_{\tau}$ and $\pi^+ \pi^{0} \bar{\nu}_{\tau}$. The $\pi^+ \pi^{0}$ is restricted to the $\rho^+$ mass region, and for brevity we also denote it as $\rho^+_{\pi^+ \pi^{0}} \bar{\nu}_{\tau}$.
We use $\tau_{e}^+$, $\tau_{\mu}^+$, $\tau_{\pi}^+$ and $\tau_{\rho}^+$ to represent the four $\tau^+$ decay modes, as well as $\tau^+_{\ell}$ $(\ell=e,~\mu)$ and $\tau^+_{h}$ $(h=\pi, \rho)$ to summarize the corresponding modes. 
Given the very small variation of detection efficiencies and backgrounds at different energy points, the eleven ST modes from the eight data samples are combined for further DT analyses.
Since the event selection criteria are almost the same as those in Sec.~\ref{sec_st}, only the differences are described below.

\begin{figure}[htbp] 
	\centering 
%	\begin{minipage}{0.23\textwidth} 
%		\centering 
		%\includegraphics[width=\linewidth]{m_rho.eps} 
		\includegraphics[width=0.48\linewidth]{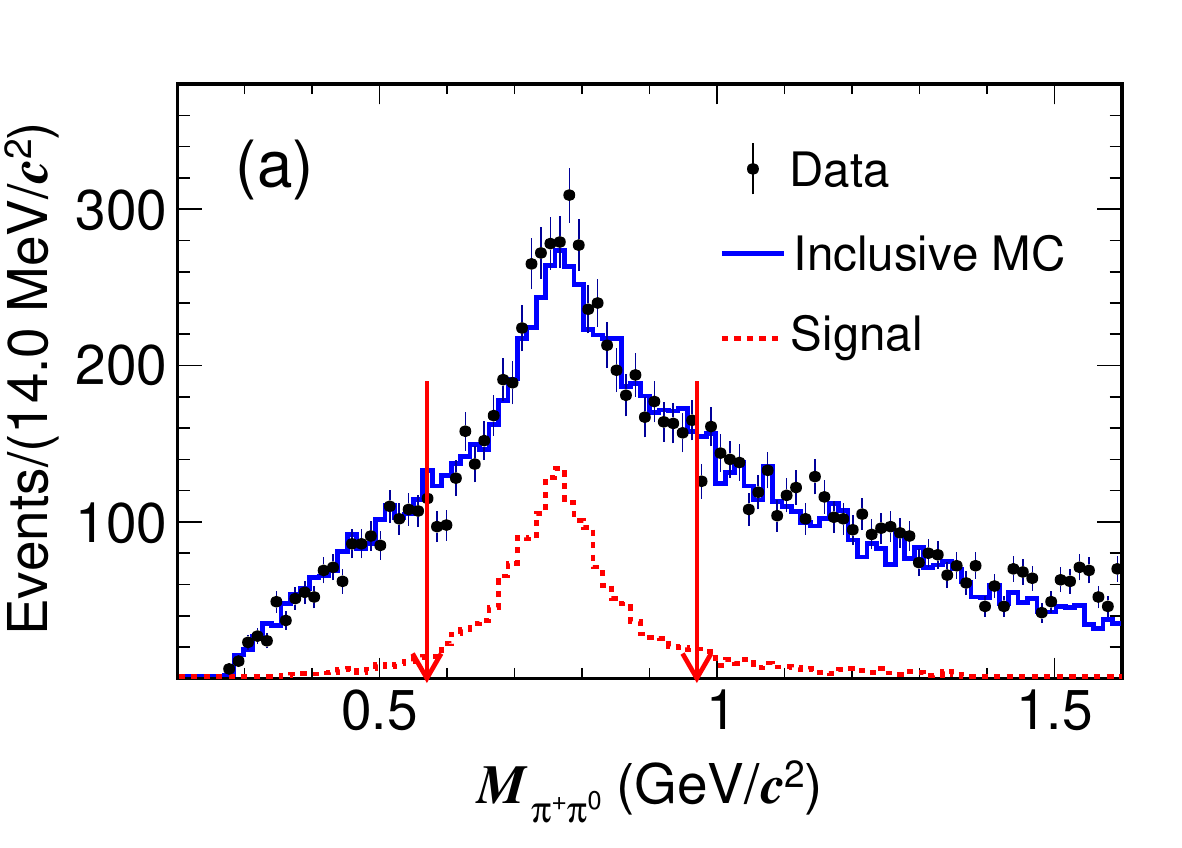} 
%		\vspace{0.2cm} {\small (a)}
%	 \end{minipage}
%	 \hfill 
%	 \begin{minipage}{0.23\textwidth}
%	 	 \centering
	 	 % \includegraphics[width=\linewidth]{mm2st.eps} 
	 	  \includegraphics[width=0.48\linewidth]{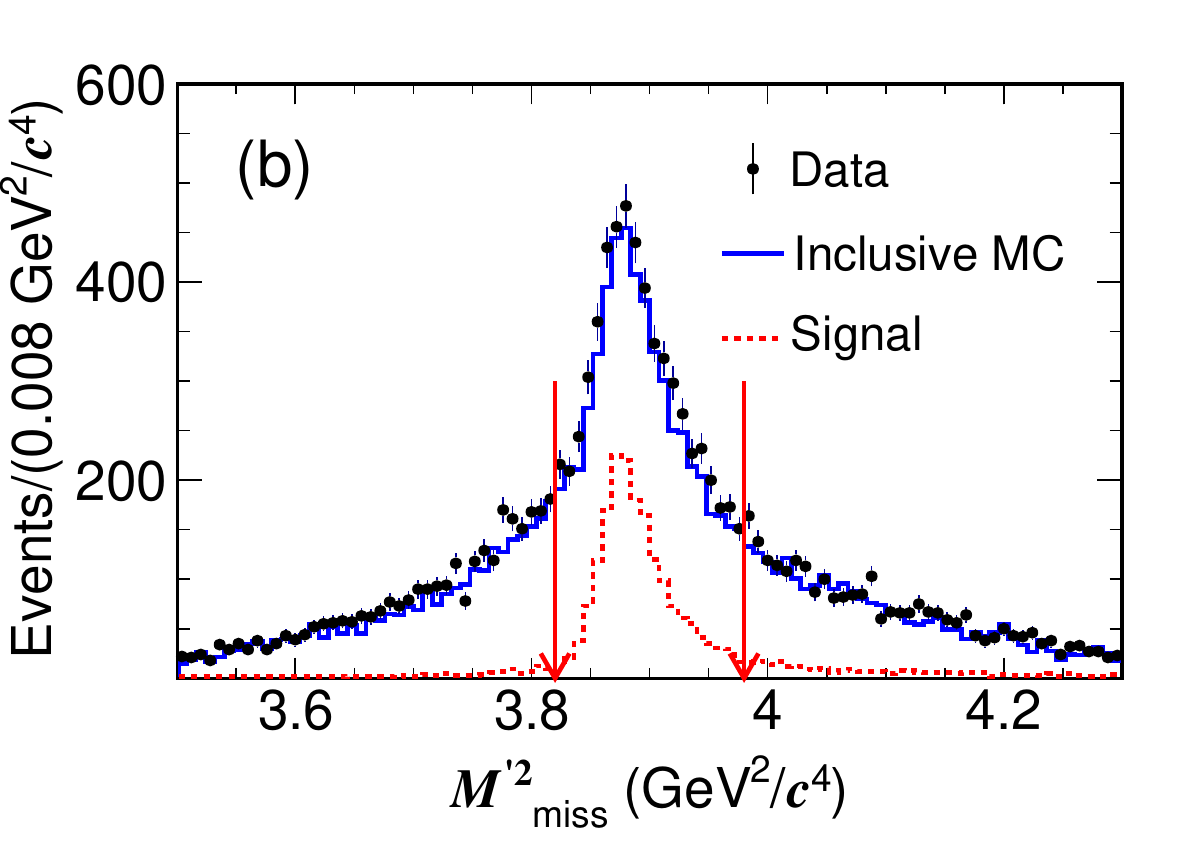} 
%	 	  \vspace{0.2cm} {\small (b)}
% 	   \end{minipage} 
    \caption{ The distributions of $M_{\pi^+\pi^0}$ (a) and
    	$M^{\prime 2}_{\rm miss}$ (b) for the $\tau^+\to \pi^+ \pi^0 \bar{\nu}_{\tau}$ decay in data and MC-simulated samples.
%    	The dots with error bars are data, 
%    	the solid blue line is the inclusive MC samples normalized to data, 
%    	and the dashed red line is the signal shape scaled to data. 
    	The area between two red arrows indicates the signal region. } 
    \label{fig:rho_mm2}
 \end{figure}

%event selection criteria for the signal decay
Exactly one additional good track, not used in the ST $D_s^-$ reconstruction and with charge opposite to the ST $D_s^-$, is required. 
%tau -> e nu nu case 
The charged track is identified as a positron if $\mathcal{L}(e)>0.001$, and $\mathcal{L}(e)/[\mathcal{L}(e)+\mathcal{L}(\pi)+\mathcal{L}(K)]>0.8$, where the likelihood $\mathcal{L}$ is estimated based on the information in the MDC, TOF and EMC.
The momentum of the positron is required to be larger than 0.2~GeV/$c$, and the ratio, $E_{\rm EMC}/p$, of the deposited energy in the EMC ($E_{\rm EMC}$) divided by the MDC momentum, must satisfy $E_{\rm EMC}/p > 0.8$.
%tau -> mu nu nu case 
The muon candidates must satisfy the requirements of $\mathcal{L}(\mu)>0.001$, $\mathcal{L}(\mu)> \mathcal{L}(K)$, $\mathcal{L}(\mu) > \mathcal{L}(e)$, and $E_{\rm EMC}\in [0.1, 0.3]$~GeV.
High-momentum muons have significant penetration depths in the MUC. To exploit this, the muon momentum is required to be larger than 0.5~GeV/$c$, and momentum- and angle-dependent requirements on the hit depth are also required, as listed in Table~\ref{tab:muonid}.
% tau -> pi nu case 
For the $\tau^+_{\pi}$ case, there must be no $\pi^0$ candidate to suppress contamination from $\tau^+_{\rho}$ decays, and the $\pi^+$ candidate must have $E_{\rm EMC}/p < 0.9$. 
To suppress background events from $D_s^+ \to \pi^+ \pi^0 \eta$ and other $\tau^+$ decays, the total energy of the residual good photons must be less than 0.3~GeV, and the cosine of the polar angle of the missing momentum must satisfy $\cos\theta_{\rm miss} < 0.9$.
%tau -> pi+ pi0 nu case 
For the $\tau^+_{\rho}$ case, the $\pi^0$ candidate is reconstructed from a pair of good photons. 
In the case of multiple $\pi^0$ candidates, the one with the minimum $\chi^2$ of the kinematic fit is chosen. 
To suppress the combinatorial backgrounds, the total energy of additional good photons must be less than 0.1~GeV, $M_{\pi^+ \pi^0}$ must lie in the $\rho^+$ signal region, i.e., $|M_{\pi^+ \pi^0} - m_{\rho^+}|<0.2$~GeV/$c^2$, and the missing mass squared $M^{\prime 2}_{\rm miss} = E_{\rm miss}^{\prime 2} - |\vec{p}^{\;\prime}_{\rm miss}|^{2}$ must be within the range of [3.82, 3.98]~GeV$^2$/$c^4$. 
Figure~\ref{fig:rho_mm2} shows the distributions of $M_{\pi^+ \pi^0}$ and $M^{\prime 2}_{\rm miss}$.
%Figure~\ref{fig:rho_mm2} shows the distributions of the $\rho^+$ candidate invariant mass and $M_{\rm miss}^{\prime,2}$ for data together with the corresponding inclusive MC and signal MC samples. The arrows indicate the selection requirements.

%the hit depth for muon candidate.
\begin{table}[hbtp]
	\begin{center}
		\caption{The requirements of the MUC hit depth, $d_{\mu^+}$, as a function of the cosine of the muon candidate angle, $\cos\theta_{\mu^+}$, and momentum, $|\vec{p}_{\mu^+}|$.}
		\setlength{\extrarowheight}{0.2ex}%the space left in the top
		\setlength{\tabcolsep}{5.0pt}% the space between two columns
		\renewcommand{\arraystretch}{1.5}
		\begin{tabular}{ccl} \hline\hline
			%		\begin{tabular}{ccl} \hline\hline
				$|\cos\theta_{\mu^+}|$ & $|\vec{p}_{\mu^+}|$ (GeV/$c$) & $d_{\mu^+}$ (cm) \\ \hline
				\multirow{5}{*}{[0.0, 0.2)}  &$(0.50, 0.61)$  &$>3$        \\
				&$(0.61, 0.75)$  &$>100 \, p_{\mu^+}-58$ \\
				&$(0.75, 0.88)$  &$>17$ \\ 
				&$(0.88, 1.04)$  &$>100 \, p_{\mu^+}-71$ \\ 
				&$(1.04, 1.20)$  &$>33$ \\ \hline
				\multirow{3}{*}{[0.2, 0.4)}  &$(0.50, 0.64)$  &$>3$  \\
				&$(0.64, 0.78)$  &$>100 \, p_{\mu^+}-61$ \\
				&$(0.78, 0.91)$  &$>17$   \\
				&$(0.91, 1.07)$  &$>100 \, p_{\mu^+}-74$   \\
				&$(1.07, 1.20)$  &$>33$   \\\hline
				\multirow{5}{*}{[0.4, 0.6)}  &$(0.50, 0.67)$  &$>3$   \\
				&$(0.67, 0.81)$  &$>100 \, p_{\mu^+}-64$ \\
				&$(0.81, 0.94)$  &$>17$ \\ 
				&$(0.94, 1.10)$  &$>100 \, p_{\mu^+}-77$ \\
				&$(1.10, 1.20)$  &$>17$ \\ \hline
				[0.60, 0.93]                 &$(0.50, 1.20)$  &$>9$ \\ \hline\hline
				%			[0.80, 0.93]                 &$[0.50, 1.20]$  &$>9$ \\ \hline\hline
			\end{tabular}
			\label{tab:muonid}
		\end{center}
	\end{table}

\begin{figure*}[htbp]\centering
		\includegraphics[width=1.0\textwidth]{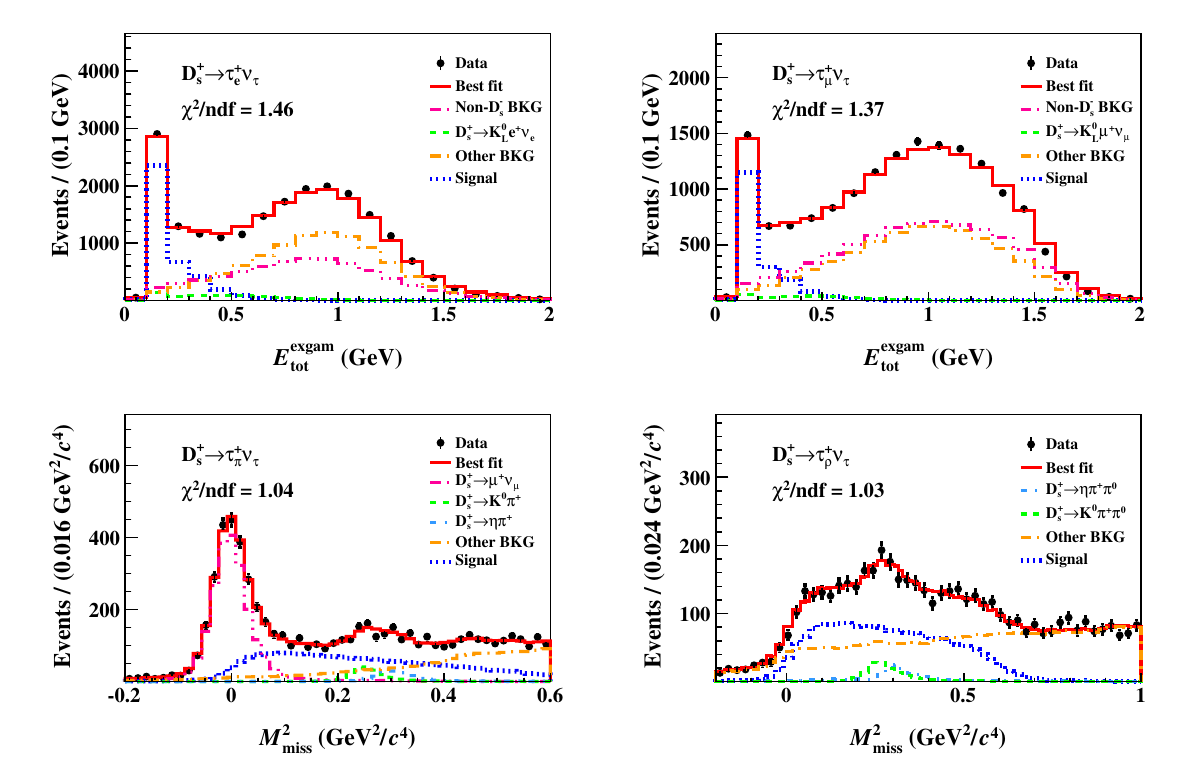}
		\caption{Simultaneous fits to the $E^{\rm exgam}_{\rm tot}$ and $M^2_{\rm miss}$ distributions for $\dstotaunuz$ candidates reconstructed in the $\tau^+_e$, $\tau^+_{\mu}$, $\tau^+_{\pi}$, and $\tau^+_{\rho}$ modes, with a common BF of ${D_{s}^{+} \to \tau^{+} \nu_{\tau}}$. The dots with error bars are data, the solid red lines are the best fits, the dotted blue lines are the signals, and the various other lines denote different backgrounds as listed.}
		\label{data_result}

	\end{figure*}

	%introduce the important fitting variables  
	Two important variables, $E_{\rm tot}^{\rm exgam}$ and $M^2_{\rm miss}$, are adopted to effectively discriminate signal from background, where the former is for the $\tau^+_{\ell}$ cases, and the latter one is for the $\tau^+_{h}$ cases.
	The $E_{\rm tot}^{\rm exgam}$ represents the total energy of good photons excluding those used in ST $D_s^-$ reconstruction, which will peak around 150~MeV due to the transition $\gamma(\pi^0)$ from $D_s^{*+}$ decays.
	The missing mass squared, $M^2_{\rm miss}$, is defined as
	\begin{equation}
		\label{eq:new_mm}
		\begin{aligned} % for multiple equations.
			M^2_{\rm miss} &= E^2_{\rm miss} - |\vec{p}_{\rm miss}|^2,\\
			E^2_{\rm miss} &= E_{\rm CM} - E_{\rm ST} - E_{\gamma(\pi^0)} - E_{\rm sig}, \\
			\vec{p}_{\rm miss} &= -\vec{p}_{\rm ST} - \vec{p}_{\gamma(\pi^0)} - \vec{p}_{\rm sig},
		\end{aligned}
	\end{equation}
	where $E_{\rm sig}$ and $\vec{p}_{\rm sig}$ are the energy and three-momentum of $\pi^+$ or $\pi^+ \pi^0$ from the $\tau^+$ decays. 
	Figure~\ref{data_result} shows the distributions of $E_{\rm tot}^{\rm exgam}$ and $M^2_{\rm miss}$ of the accepted candidates for $D_s^+\to \tau^+\nu_\tau$ in data.

	%background analysis
	%2025.3.24 18:24, revised here.
	According to the MC studies, the potential background events for the $\tau^+_{\ell}$ case can be categorized into three sources.
	The first is the non-$D_s^-$ background where the ST $D_s^-$ is incorrectly reconstructed. The second one is the peaking background $D_s^+ \to K_L^0 \ell^+ \nu_{\ell}$ along with the correctly reconstructed ST $D_s^-$, which mimics the signal when the $K_L^0$ leaves little energy in the detector.
	The third one includes the other backgrounds, and is dominated by $D_s^+ \to X e^+ \nu_e$ decays with $X = \eta$, $\eta^{\prime}$, $\phi$, $f_0(980)$ for the $\tau^+_e$ case, and $D_s^+ \to \eta \mu^+ \nu_\mu$ and $D_s^+ \to \eta \pi^+ \pi^0$  for the $\tau^+_\mu$ case.
	%2025.3.24, revised here
	%add Ds+ -> eta pi+ pi0 for tau -> mu nu nu case.
	For the $\tau^+_{\pi}$ case, the peaking background is mainly from $D_s^+ \to \mu^+ \nu_{\mu}$, since the muon has a similar mass as pion.
	The $D_s^+ \to K^0 \pi^+$ and $D_s^+ \to \eta \pi^+$ backgrounds form a peak around $M^2_{\rm miss}$ close to the $K^0$ and $\eta$ masses, respectively. 
	For the $\tau^+_{\rho}$ case, the peaking backgrounds from $D_s^+ \to K^0 \pi^+ \pi^0$ and $D_s^+ \to \eta \pi^+ \pi^0$ remain due to the missing $K^0$ and $\eta$. For both of these $\tau^+_{h}$ cases, the remaining backgrounds are mostly combinatorial backgrounds. 
	
	%\section{Simultaneous fit} 
	%\subsection{Branching fraction of $D^+_s\to \tau^+\nu_\tau$}
	
	We perform a simultaneous fit to the $E_{\rm tot}^{\rm exgam}$ distributions for the $\tau^+_{\ell}$ cases and $M^2_{\rm miss}$ for the $\tau^+_{h}$ cases, where the DT yield in each case is constrained to share the same BF of $D^+_s\to \tau^+\nu_\tau$ according to Eq.~(\ref{eq1_br_cal}).
	In the fit, the signal shapes are modeled by the MC-simulated shapes. 
	For the $\tau^+_\ell$ cases, the non-$D_s^-$ background is described by the events in the $M_{\rm ST}$ sideband regions ([1.895, 1.920] and [2.010, 2.035]~GeV/$c^2$) after requiring the DT selections, and its size is fixed in accordance with the ratio of the yields in the $M_{\rm ST}$ signal region over those in the sideband regions, where the yields are obtained by fitting to the $M_{\rm ST}$ distribution.
	The shape of the peaking background of $D_s^+ \to K_L^0 \ell^+ \nu_{\ell}$ is described by the MC-derived shape corrected with a two-dimensional (polar angle and momentum) data-MC difference for the $K_L^0$ detector response, in order to reduce the difference of the $K_L^0$ reconstruction efficiencies between data and MC simulation. 
	The polar angle and momentum-dependent correction factors are extracted from a control sample of $D^0 \to K_L^0 \pi^+ \pi^-$. 
	The size is fixed according to the $\mathcal{B}({D_s^+ \to K^0 e^+ \nu_{e}})= (2.98\pm0.23\pm0.12)\times10^{-3}$ measured by BESIII~\cite{lilei_paper_Ds_k0enu}.
	%The shape of the other backgrounds is extracted from the MC simulations, and the size is left free. 
	For the $\tau^+_h$ cases, the yield for each peaking background is floated due to their distinct shapes in the $M^2_{\rm miss}$ spectrum, except for $D_s^+ \to \eta \pi^+ \pi^0$ which has a fixed size estimated via the PDG BF~\cite{pdg}. 
	The shape of the other backgrounds is extracted from the MC simulations, and the size is left free. 
	The simultaneous fit results are shown in Fig.~\ref{data_result}. The DT efficiencies are obtained by analyzing the inclusive MC sample.
	The BF of $\dstotaunuz$ is determined to be $(5.37 \pm 0.08)\%$, where the uncertainty is statistical only. This BF corresponds to a total DT yield of $9851\pm146$.

	\section{Systematic uncertainties}
	  
	%\iffalse
	In the BF measurement, the systematic uncertainties arising from ST selection criteria are mostly canceled, except for those from the background fluctuations in the fitted ST yields and the tag bias.
The total ST yield of all eleven ST modes is $(801.7\pm3.8)\times10^{3}$, 
resulting in the statistical uncertainty of $0.5\%$. Here, we only consider the statistical fluctuation related to the background of the ST side, as it is not correlated with the DT sample directly. Hence, a systematic uncertainty of 0.5\% is assigned. 
%	The systematic uncertainty from the background fluctuations, 
%	mainly resulting in the statistical uncertainty of the total ST yield $((801.7\pm3.8)\times10^{3}$), 
%	is estimated to be $\sqrt{3800*3800-801700}/801700=0.5\%$. 
	MC studies show that there exists a difference in the ST efficiencies between the inclusive MC and signal MC samples, which is called the tag bias. 
	The difference weighted by the ST yields in various data samples, 0.3\%, is taken as the systematic uncertainty from the tag bias.

	The systematic uncertainty from the photon detection efficiency is studied with the control sample of $J/\psi \to \pi^+ \pi^- \pi^0$, which is determined to be 0.5\% per photon in the EMC barrel region, and 1.5\% per photon in the EMC end cap regions~\cite{gamma}. After weighted according to the spatial distribution of photons in the signal MC sample of $D_s^{*+} \to \gamma D_s^{*+}$, the systematic uncertainty from the photon detection is 0.52\%.
	The systematic uncertainty from the $\pi^0$ reconstruction is studied with the control sample of $e^+ e^- \to K^+ K^- \pi^+ \pi^- \pi^0$, and is assigned to be 2.0\%~\cite{pi0_reconstruction_eff}. After re-weighting by the BFs of $D_s^{*+} \to \gamma D_s^+$ and $D_s^{*+}\to \pi^0 D_s^+$~\cite{pdg}, the systematic uncertainty from the transition $\gamma,\pi^0$ reconstruction is estimated to be 0.6\%, and the systematic uncertainty from the $\pi^0$ reconstruction for the $\tau^+_{\rho}$ case is 0.2\%~\cite{tauvyue}.

	Utilizing the control samples of $e^+ e^- \to \gamma \ell^+ \ell^-$, $e^+ e^- \to K^+ K^- \pi^+ \pi^- (\pi^0)$ and $e^+ e^- \to \pi^+ \pi^- \pi^+ \pi^- (\pi^0)$, 
	the tracking and PID efficiencies for the $e^+$, $\mu^+$, and $\pi^+$ are investigated in different polar angle and momentum intervals, and the efficiencies are then re-weighted with the corresponding events in the signal MC samples.  
	The differences in the re-weighted tracking (PID) efficiencies between data and MC simulation are $(-0.2\pm0.1)\%$, $(-0.3\pm0.2)\%$, $(-0.3\pm0.3)\%$ and $(-0.1\pm0.2)\%$ for the tracking of $e^+$, $\mu^+$,  $\pi^+$(from $\tau_{\pi}^+$ case) and $\pi^+$(from $\tau_{\rho}^+$ case), respectively, while we obtain $(-0.3\pm0.1)\%$, $(-12.0\pm0.3)\%$, $(-1.1\pm0.3)\%$ and $(-0.4\pm0.1)\%$ for the PID of $e^+$, $\mu^+$, $\pi^+$(from $\tau_{\pi}^+$ case) and $\pi^+$(from $\tau_{\rho}^+$ case), respectively.
	The mean values are then used to correct the detection efficiency. 
	The residual uncertainties, weighted by individual $\tau^+$ decay BFs, are taken as the corresponding systematic uncertainties; these are all equal or rounded up to 0.1\% for tracking or PID per $e^+$, $\mu^+$, or $\pi^+$. The total systematic uncertainties from tracking and PID are both 0.2\% after adding them in quadrature.

	The efficiencies for the requirements of no other charged tracks and the total energy of extra photons for the $\tau^+_h$ cases, are investigated with the DT sample of $D_s^+ \to \eta \pi^+$. The data-MC difference in the acceptance efficiencies, 2.2\%, is treated as the corresponding systematic uncertainty.
	The requirement of $M^{\prime\, 2}_{\rm miss}$ is altered by additionally smearing a Gaussian function, whose parameters are obtained by fitting the data distribution. 
	The difference in the signal efficiency, 0.8\%, is taken as the systematic uncertainty from this requirement.
	The efficiencies for the requirements on $\cos\theta_{\rm miss}$, and the $E_{\rm EMC}/p$ ratio, are studied by the DT samples of $D_s^+ \to K^+ K^- \pi^+$ and $D_s^+ \to K_S^0 \pi^+$. The data-MC differences of the acceptance efficiencies, 1.1\% and 0.2\%, are taken as individual systematic uncertainties.
	The sum in quadrature of each item after weighting by the four $\tau^+$ decay BFs, 0.5\%, is assigned as the systematic uncertainty from the requirements on the DT side.

	The systematic uncertainties in the simultaneous fit arise from the signal shapes and the background shapes. 
	Alternative fits are performed with the signal shape described by the MC simulated shape convolved with a Gaussian function for the $\tau^+_{\ell}$ cases, while the MC simulated shape for the $\tau^+_{h}$ cases. 
	The sum in quadrature of the relative changes on the measured BFs, 0.5\%, is taken as the systematic uncertainty from the signal shape. 
	For the background shapes, the non-$D_s^-$ background shape is replaced with the events in the $M_{\rm ST}$ signal region from the inclusive MC sample, the peaking background shapes are altered by the MC simulated shape convolved with a Gaussian function, and the other background shape is extracted by varying the relative fraction of the dominant background component by $\pm 30\%$ for the $\tau^+_{\ell}$ cases and $\pm 1\sigma$ according to the BF for the $\tau^+_h$ cases.
	The sum in quadrature of the relative changes on the measured BF in each term, 0.2\%, is used as the systematic uncertainty from the background shapes. 
	The uncertainties from fixed sizes of peaking backgrounds are obtained by sampling their values from Gaussian distributions based on the means and uncertainties of their BF $10^4$ times. 
	The relative change on the measured BF is then fitted by a Gaussian function. 
	The quadratic sum of their fitted resolutions, 0.3\%, is assigned as the systematic uncertainty from the fixed sizes of peaking backgrounds. 
	Adding these three contributions in quadrature, the systematic uncertainty in the simultaneous fit is determined to be 0.6\%.

	The systematic uncertainties of the quoted BFs for the $\tau^+_e$, $\tau^+_\mu$, $\tau^+_\pi$, and $\tau^+_{\rho}$ decays~\cite{pdg} are 0.23\%, 0.23\%, 0.46\%, and 0.35\%, respectively. 
	After weighting by the individual signal yields, the systematic uncertainty from the quoted BFs is found to be 0.2\%. 
	The uncertainty from the limited MC statistics is assigned to be 0.2\%. 
	
	%\subsection{Summary of the systematic uncertainty sources}
	
	%The above systematic uncertainties are categorized into the $\tau^+$-decay correlated and uncorrelated sources, as summarized in Table~\ref{summary_sys_all_ds_taunu}.
	%The uncorrelated ones consist of those from ST yield, tag bias, and transition $\gamma,\pi^0$ reconstruction, while the rest are correlated ones. 
	%The total systematic uncertainty obtained by the quadrature of the contributions is 1.2\%.
	  Table~\ref{summary_sys_all_ds_taunu} summarizes the above systematic uncertainties in the BF measurement of $D_{s}^{+} \to \tau^{+} \nu_{\tau}$. The total systematic uncertainty is estimated to be 1.2\% by summing them in quadrature.
	\begin{table}[htbp]
		\begin{center}
			\caption{Summary of the relative systematic uncertainties in the BF measurement of $D_{s}^{+} \to \tau^{+} \nu_{\tau}$, in \%.}
			\label{summary_sys_all_ds_taunu}
			%	\begin{tabular}{l|r}\hline\hline
				\setlength{\extrarowheight}{0.2ex}%the space left in the top
				\setlength{\tabcolsep}{5.0pt}% the space between two columns
				\renewcommand{\arraystretch}{1.5}
				\begin{tabular}{lc}
                    \hline\hline
					Source & Uncertainty\\\hline
					ST yield & 0.5\\ 
					Tag bias & 0.3\\
					Transition $\gamma,\pi^0$  & 0.6\\
					Tracking   & 0.2\\
					PID        & 0.2 \\
					Other DT requirements     & 0.5\\
					Simultaneous fit & 0.6\\ 
					$\pi^0$ from $\rho^+$ decays & 0.2\\
					Quoted BFs       & 0.2\\
					MC statistics    & 0.2\\\hline
					Total            & 1.2 \\ %2024.5.15
					\hline\hline
				\end{tabular}
			\end{center}
		\end{table}
		
		%\fi
		
		\section{RESULTS}
		
		%\iffalse
		
		Taking the systematic uncertainties into account, the BF of $D_s^+ \to \tau^+ \nu_{\tau}$ is determined to be $\mathcal{B}({D_s^+ \to \tau^+ \nu_{\tau}}) = (5.37 \pm 0.08_{\rm stat} \pm 0.06_{\rm syst})\%$. 
		To determine $f_{D_s^+} |V_{cs}|$, radiative corrections for $\mathcal{B}({D_s^+ \to \tau^+ \nu_{\tau}})$ are necessary as stated by Ref.~\cite{shortEM1}, 
while radiative corrections associated with the subsequent $\tau^+$ decays are not required~\cite{L3:2001szz}. 
For $D_s^+ \to \tau^+ \nu_{\tau}$ process, the effect of $D_s^+ \to \gamma \tau^+ \nu_{\tau}$ can be neglected, 
as there is no significant helicity suppression in the $D_s^+ \to \tau^+ \nu_{\tau}$ decay, 
rendering the radiative corrections induced by such processes insignificant~\cite{EM3}. 
The known short-distance electroweak correction increases $\mathcal{B}({D_s^+ \to \tau^+ \nu_{\tau}})$ by 1.8\%~\cite{shortEM1,shortEM2}, 
and the long-distance electroweak correction lowers $\mathcal{B}({D_s^+ \to \tau^+ \nu_{\tau}})$ by 2.5\%~\cite{longEM1}; assign a 0.6\% uncertainty to the total. 
Combining our BF with the world average values of $G_F,m_{D_s^+},m_{\tau^+}$ and the $D_s^+$ lifetime $\tau_{D_s^{+}}$ into Eq.~(\ref{decayratedstolnu_draft}) and incorporating the above corrections, 
the product of $f_{D_s^+}$ and $|V_{cs}|$ is estimated to be $f_{D_s^+} |V_{cs}| = (248.2 \pm 1.9_{\rm stat} \pm 1.4_{\rm syst} \pm 0.6_{\rm input} \pm 0.8_{\rm EM})$~MeV, 
where the third uncertainty primarily stems from $\tau_{D_s^{+}}$, while the fourth uncertainty originates from radiative corrections. 
Using the $|V_{cs}| = 0.97349 \pm 0.00016$ from the CKMfitter group~\cite{pdg}, 
the $f_{D_s^+}$ is extracted to be $f_{D_s^+} = (255.0 \pm 1.9_{\rm stat} \pm 1.4_{\rm syst} \pm 0.6_{\rm input} \pm 0.8_{\rm EM})~\mathrm{MeV}$, 
which is $2.0\sigma$ higher than the latest LQCD average value~\cite{fds_flag} and $1.6\sigma$ lower than the QCD sum rule prediction~\cite{qcdsumrules}. 
On the other hand, taking the value $f_{D_s^+} = (249.9 \pm 0.5)$~MeV from LQCD calculations~\cite{fds_flag} as an input, 
the $|V_{cs}|$ is obtained to be $|V_{cs}| = 0.993 \pm 0.008_{\rm stat} \pm 0.006_{\rm syst} \pm 0.003_{\rm input} \pm 0.003_{\rm EM}$, which is about $1.8\sigma$ higher than the value from the CKMfitter group~\cite{pdg}. 
The current experimental uncertainties remain significantly larger than the uncertainties of the LQCD calculations and the CKMfitter global fit. 
Therefore, it is necessary to further improve the experimental precision with a large statistics in the future, 
in order to enable more stringent tests of the SM and search for new physics beyond the SM. 
%Furthermore, the measured product $f_{D_s^+}|V_{cs}|$ can also be combined with complementary experimental and theoretical inputs to provide additional consistency tests of the SM. 
Combining the obtained $f_{D_s^+} |V_{cs}|$ with its counterpart $f_{D^+} |V_{cd}|$ measured by BESIII~\cite{munutengjiao}, 
with $|V_{cd}/V_{cs}| = 0.2310 \pm 0.0007$ from the SM global fit~\cite{pdg}, 
gives $f_{D_s^+}/f_{D^+} = 1.194 \pm 0.015_{\rm stat} \pm 0.009_{\rm syst} \pm 0.005_{\rm input}$. 
This result is consistent with the predictions from LQCD~\cite{fds_flag} and QCD sum rules~\cite{qcdsumrules} within 1$\sigma$. 
Alternatively, using the LQCD input $f_{D_s^+}/f_{D^+} = 1.1788 \pm 0.0046$~\cite{fds_flag}, 
we obtain $|V_{cd}/V_{cs}| = 0.228 \pm 0.003_{\rm stat} \pm 0.002_{\rm syst} \pm 0.001_{\rm input}$, 
which agrees with the expectation based on the values of $|V_{cs}|$ and $|V_{cd}|$ from the CKMfitter group~\cite{pdg} within 1$\sigma$. Here, only the systematic uncertainty associated with the radiative correction is canceled, since the two data samples were collected in different years.
		
		The average BF for the $D_s^+\to\tau^+\nu_\tau$ decay is determined to be $(5.41 \pm 0.08)\%$ by a re-weighted analysis combining this work with previous measurements reported by the CLEO~\cite{cleo_dstotaunu_enunu2,cleodstotaunu2,cleodstotaunu3}, $BABAR$~\cite{babardstotaunu}, Belle~\cite{belldstotaunu}, and BESIII~\cite{bes4009,tauvwuxiao} collaborations. By utilizing the world average value of $\mathcal{B}({{D_s^+ \to \mu^+ \nu_\mu}})$~\cite{pdg} and the results presented in this work, the ratio of the BFs is determined to be $\mathcal{B}({{D_s^+ \to \tau^+ \nu_\tau}}) / \mathcal{B}({{D_s^+ \to \mu^+ \nu_\mu}}) = 10.04 \pm 0.29$, which is consistent with the SM prediction $9.75 \pm 0.01$.
		
		\section{SUMMARY}
		
		Based on 7.33~fb$^{-1}$ of $e^+e^-$ collision data collected at $E_{\rm CM}=$4.128-4.226~GeV, we report the absolute BF measurement of $D_s^+ \to \tau^+ \nu_{\tau}$ with a simultaneous fit to four $\tau^+$ decay modes. The measured branching fraction is $\mathcal{B}({D_s^+ \to \tau^+ \nu_{\tau}}) = (5.37 \pm 0.08_{\rm stat} \pm 0.06_{\rm syst})$\%. The precision is improved by a factor of 1.5 compared to the previous best individual measurement~\cite{tauvhuijing}, and by a factor of 2.6 compared to previous BESIII result obtained with the same fit method but a different data sample~\cite{tauvwuxiao}.
		Using as inputs $f_{D_s^+}$ from LQCD or $|V_{cs}|$ from the SM global CKM fit, we obtain $|V_{cs}| = 0.993 \pm 0.008_{\rm stat} \pm 0.006_{\rm syst} \pm 0.003_{\rm input} \pm 0.003_{\rm EM}$, and $f_{D_s^+} = (255.0 \pm 1.9_{\rm stat} \pm 1.4_{\rm syst} \pm 0.6_{\rm input} \pm 0.8_{\rm EM})$~MeV, respectively, which are the most precise determinations of these observables to date.
		Combining these results with the world average value of $\mathcal{B}(D_s^+ \to \mu^+ \nu_{\mu})$, the test of LFU between $\tau$ and $\mu$ leptons gives a ratio of $\mathcal{B}({{D_s^+ \to \tau^+ \nu_\tau}}) / \mathcal{B}({{D_s^+ \to \mu^+ \nu_\mu}}) = 10.04 \pm 0.29$, consistent with the SM expectation.

		\section*{\boldmath ACKNOWLEDGMENTS}
		
		%\input{acknowledgement_2025-09-10.tex}

		%\iffalse
		The BESIII Collaboration thanks the staff of BEPCII (https://cstr.cn/31109.02.BEPC) and the IHEP computing center for their strong support.  
		This work is supported in part by National Key R\&D Program of China under Contracts No. 2023YFA1606000; Outstanding Youth Foundation of Henan Province No. 252300421219; 
		National Natural Science Foundation of China (NSFC) under Contracts Nos. 12305102, 12275067, 12275068, 12105077, 11635010, 11935015, 11935016, 11935018, 12025502, 12035009, 12035013, 12061131003, 12192260, 12192261, 12192262, 12192263, 12192264, 12192265, 12221005, 12225509, 12235017, 12361141819; 
		the Chinese Academy of Sciences (CAS) Large-Scale Scientific Facility Program; CAS under Contract No. YSBR-101; 
		100 Talents Program of CAS; The Institute of Nuclear and Particle Physics (INPAC) and Shanghai Key Laboratory for Particle Physics and Cosmology; 
		Science and Technology R$\&$D Program Joint Fund Project of Henan Province  (Grant No.225200810030), Natural Science Foundation of Henan Province (Grant Nos. 252300421491, 242300420250); Science and Technology Innovation Leading Talent Support Program of Henan Province; 
		Agencia Nacional de Investigación y Desarrollo de Chile (ANID), Chile under Contract No. ANID PIA/APOYO AFB230003; ERC under Contract No. 758462; German Research Foundation DFG under Contract No. FOR5327; Istituto Nazionale di Fisica Nucleare, Italy; Knut and Alice Wallenberg Foundation under Contracts Nos. 2021.0174, 2021.0299; Ministry of Development of Turkey under Contract No. DPT2006K-120470; National Research Foundation of Korea under Contract No. NRF-2022R1A2C1092335; National Science and Technology fund of Mongolia; Polish National Science Centre under Contract No. 2024/53/B/ST2/00975; STFC (United Kingdom); Swedish Research Council under Contract No. 2019.04595; U. S. Department of Energy under Contract No. DE-FG02-05ER41374.

		%\fi

		%\fi
		
	\end{document}